# Diffusion in binary and pseudo-binary L1$_2$ indides, stannides, gallides and aluminides of rare-earth elements as studied using perturbed angular correlation of [111]In/Cd

Randal Newhouse[a], Justine Minish[b] and Gary S. Collins[c]

Department of Physics and Astronomy, Washington State University, Pullman, WA, USA

[a] randynewhouse@gmail.com, [b] 12jmmini@alma.edu, [c] collins@wsu.edu

**Abstract.** Diffusional jumps can produce fluctuating electric field gradients at nuclei of jumping atoms. Using perturbed angular correlation of gamma rays (PAC), jumps of probe atoms cause nuclear quadrupole relaxation that can be fitted to obtain the mean jump frequency. An overview is given of the application of this approach to highly ordered intermetallic compounds having the L1$_2$ (Cu$_3$Au) crystal structure. New results are then presented for jump frequencies of [111]In/Cd probe atoms in pseudo-binary L1$_2$ compounds of the forms In$_3$(La$_{1-x}$Pr$_x$) and (In$_{1-x}$Sn$_x$)$_3$La. For the mixed rare-earth system, jump frequencies are found to scale with composition between jump frequencies of the end-member phases In$_3$La and In$_3$Pr. However, for the mixed sp-element system, a large decrease in jump frequency is observed as Sn atoms substitute for In-atoms. This difference in behavior appears to depend on whether atomic disorder is on the diffusion sublattice (In-Sn substitution), as opposed to a neighboring sublattice (La-Pr substitution), whether or not there is a difference in diffusion mechanism between end-member phases, and/or whether or not there is a valence difference between the mixing atoms. All three conditions apply for only (In$_{1-x}$Sn$_x$)$_3$La.

**Introduction**

In a seminal study in 2004 on In$_3$La, it was shown that jump frequencies of PAC probe atoms could be measured when diffusional jumps occur with changes in orientation or magnitude of electric field gradients at the probe sites [1]. In$_3$La has the L1$_2$ structure A$_3$B, in which jumps on the A-sublattice lead to reorientation of the electric field gradient (EFG) along the three [100] cube directions. Such jumps produce decoherence of nuclear quadrupole precessions that is exhibited as "damping" in time-domain PAC perturbation functions [1]. In good approximation, as shown in [2], the damping time is equal to the mean residence time of a probe atom on a site, or inverse of the total jump frequency $w$. For cubic compounds, $w$ is related to the diffusivity $D$ via $D = f\ell^2 w/6$, in which $\ell$ is the jump distance and $f$ is the correlation factor for diffusion.

Subsequent measurements were made on a wide range of rare-earth indides [2, 3, 4], stannides [2,5] and gallides [6] of L1$_2$ structure. These compounds are all highly-ordered "line compounds" in binary phase diagrams, meaning that differences between the two boundary compositions of the phase are less than ~1 at.%, perhaps only 0.1 at.%. In all these measurements, the tracers were excited–state [111]Cd probe atoms starting on the A-sublattice, immediately following decay of [111]In.

Nevertheless, it was found as early as in ref. [1] that the measured jump frequencies are strongly dependent on the composition, with jump frequencies 10-100 times greater at one boundary composition than the other [7]. This difference gives insight into the dominant diffusion mechanism [4]. Vacancy concentrations vary monotonically with composition, so that, e.g., the A-vacancy concentration must increase as the composition becomes more B-rich. Consequently, if diffusion of probe atoms on the A-sublattice in A$_3$B involves only A-vacancies, then the jump frequency will be greater at the A-poor boundary composition. Therefore, the phase boundary at which one observes the greater probe jump frequency identifies the primary vacancy-type responsible for diffusion on the A-sublattice. For example, the jump frequency has been determined experimentally to be greater at the Sn-poor boundary composition in the stannide $Sn_3La$, indicating that Sn-vacancies dominate diffusion on the Sn-sublattice [5].

For lanthanide tri-indides $In_3R$, the jump frequency was found to be greater at the In-rich phase boundaries in the light lanthanide indides (R= La,Ce,Pr…), but greater at the In-poor phase boundaries in the heavy lanthanide indides (R= Lu,Tm,Er,Ho,Dy,Tb,Gd…) [4]. Thus, there is a change in the dominant diffusion mechanism along the series from one dominated by B-vacancies for the light lanthanides to one dominated by A-vacancies for the heavy lanthanides. For $In_3La$ and other light lanthanide indides, one might suppose that B-vacancies have a lower partial formation energy than A-vacancies, making the B-vacancy concentration relatively greater. However, recent WIEN2k calculations of partial defect formation energies in $In_3La$ and $In_3Lu$ show just the opposite [8]. Thus, the change in diffusion mechanism along the $In_3R$ series cannot be presently understood on the basis of vacancy concentrations. As an alternative, migration energies might vary along the series, but the calculations of migration energies are much more laborious than defect energies.

In the present work, we investigate what happens when one forms a random mixture of different elements on either of the two sublattices of the $A_3B$ structure. We chose to study $In_3(La_{1-x}Pr_x)$, in which La and Pr atoms share the rare-earth sublattice, and for which both end-member binary phases, $In_3La$ and $In_3Pr$, diffusion is dominated by R-vacancies. We also selected $(In_{1-x}Sn_x)_3La$, in which different vacancies dominate diffusion in the end-member phases.

**Experiment**

Samples of $In_3(La_{1-x}Pr_x)$, with x= 0.25, 0.50, 0.75, and $(In_{1-x}Sn_x)_3La$, with x= 0.13, 0.25, 0.50, 0.75 were made by arc-melting high-purity metal foils under argon, as in previous work. Nominal compositions were calculated from the masses of elements. Compositions of the mixed elements, defined by *x*, may differ due to mass losses during melting, and could deviate from nominal compositions by up to $\pm 0.10$. X-ray measurements on selected mixed-element samples showed only the $L1_2$ crystal structure and that lattice parameters obeyed Vegard's law in good approximation. PAC spectra also gave no indication that there was more than one phase present. Consequently, it is believed that the (La,Pr) and (In,Sn) elements mixed randomly on their respective sublattices, making two different kinds of pseudo-binary $L1_2$ phases. Measurements reported in this paper were made on samples containing a slight excess of In or (In,Sn). Measurements on samples having a slight deficit of In or (In,Sn) will be reported elsewhere.

PAC measurements were made using a four-detector spectrometer, as in previous work [9]. Fig. 1 shows spectra for the alloys and end-member phases measured at 500°C. At room temperature, static quadrupole interaction frequencies $\omega_1 \equiv \frac{3\pi}{10} eQV_{zz}/\hbar$ were 68.0, 74.0 and 31.4 Mrad/s, respectively, for $In_3La$, $In_3Pr$ and $Sn_3La$. $In_3(La_{1-x}Pr_x)$ has similar end-member frequencies, and only a slight change in the mean quadrupole interaction frequency was observed as a function of *x*. However, $(In_{1-x}Sn_x)_3La$ has a large difference in frequencies, so that the effects of replacing individual In atoms with Sn atoms are larger. Quadrupole interactions for this series exhibited significant inhomogeneous broadening, and it was not possible to analyze the *x*=0.50 sample.

Spectra were measured at temperatures up 900 °C. Most measurements were made in the slow-fluctuation regime, in which jump frequencies are less than the quadrupole interaction frequency [1,10]. In this regime, the PAC quadrupole perturbation function $G_2(t)$, which heuristically exhibits spin rotations of the nuclear moment, is a damped version of the static function $G_2^{static}(t)$:

$$G_2(t) \cong \exp(-wt) G_2^{static}(t). \tag{1}$$

For higher jump frequencies (top two spectra on left and on right in Fig. 1), the periodicity of the quadrupole precessions becomes invisible due to rapid motional averaging of EFGs among the three orthogonal cube directions, approaching an average EFG of zero in the limit of very high jump frequency.

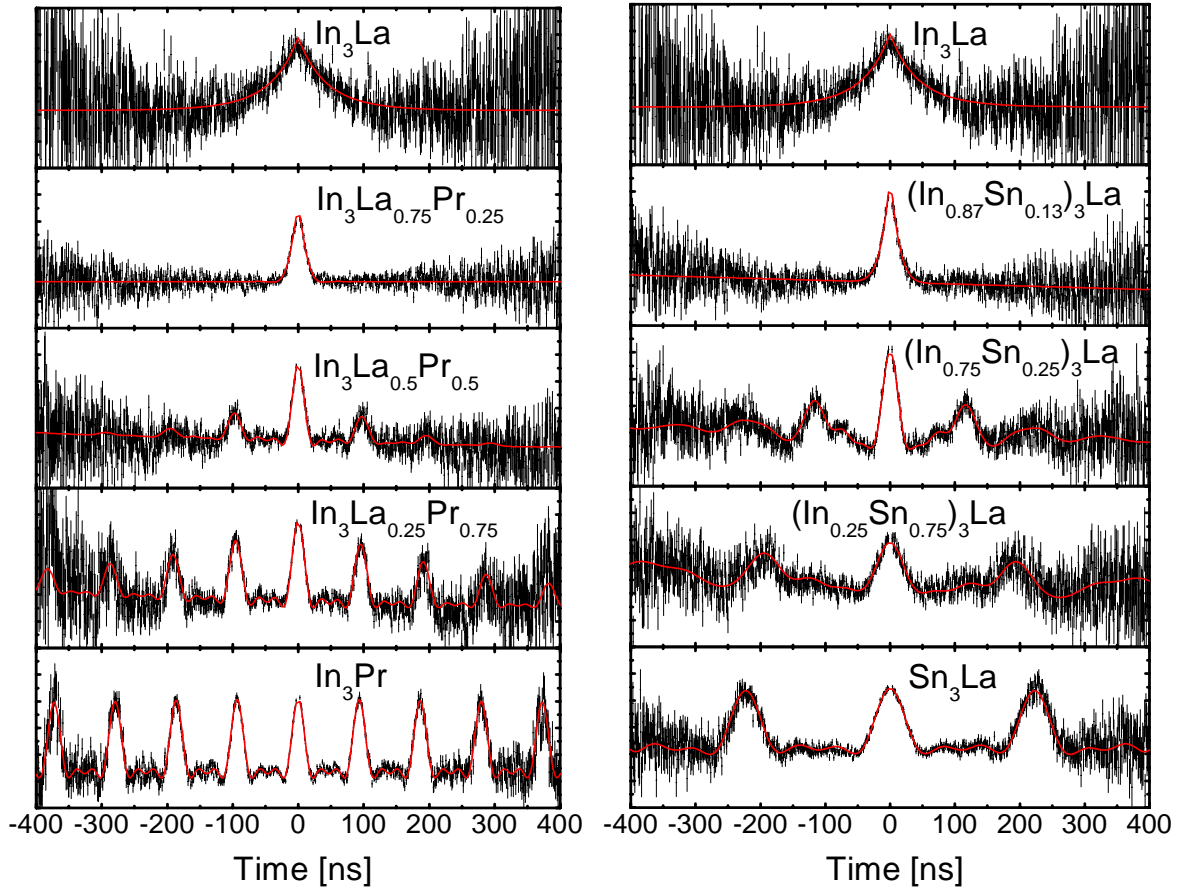

Fig. 1. PAC spectra of pseudo-binary phases of $In_3(La_{1-x}Pr_x)$, left, and $(In_{1-x}Sn_x)_3La$, right, all measured at 500 °C.

**Experimental results**

Figure 1 shows PAC spectra measured at 500°C for nominal compositions $x$= 0, 0.25, 0.50, 0.75, 1.0, of $In_3(La_{1-x}Pr_x)$, left, and for compositions $x$= 0, 0.13, 0.25, 0.75, 1.0, of $(In_{1-x}Sn_x)_3La$, right. It is evident from the figure that damping is greater for compositions approaching the end-member phase $In_3La$ in both series. For each mixed-phase sample, a spectrum was measured at room temperature, where there is no diffusional broadening, and fitted to obtain a reference $G_2^{static}(t)$ function, including inhomogeneous broadening. Subsequent measurements on a sample at elevated temperature were then fitted with Eq. 1 to obtain the mean jump frequency $w$, using the reference $G_2^{static}(t)$ for that sample. A few spectra measured in the fast fluctuation regime at high temperature were fitted using methods outlined in ref. [1].

Fig. 2 shows Arrhenius plots of fitted jump frequencies for the five samples in each series, with data for $In_3(La_{1-x}Pr_x)$ and $(In_{1-x}Sn_x)_3La$ on left and right, respectively. For reference, jump-frequency activation enthalpies for In and Sn-rich phases of the end-member compounds were 0.531(2), 1.20(3) and 1.17(4) eV, respectively, for $In_3La$, $In_3Pr$ and $Sn_3La$. It can be seen that jump frequencies are 100-1000 times greater for the end-member phase $In_3La$ than for either $In_3Pr$ or $Sn_3La$. There are major qualitative differences between results for the two pseudo-binary series. For $In_3(La_{1-x}Pr_x)$, jump frequencies decrease monotonically with increasing Pr concentration. However, for $(In_{1-x}Sn_x)_3La$, the jump frequency decreases very rapidly as In starts to be replaced with Sn. In addition, changes are not at all monotonic, with jump frequencies observed to be slightly lower for $(In_{0.25}Sn_{0.75})_3La$ than for $Sn_3La$. These observations motivated a search for microscopic models to simulate the observed composition dependences that are discussed in the following.

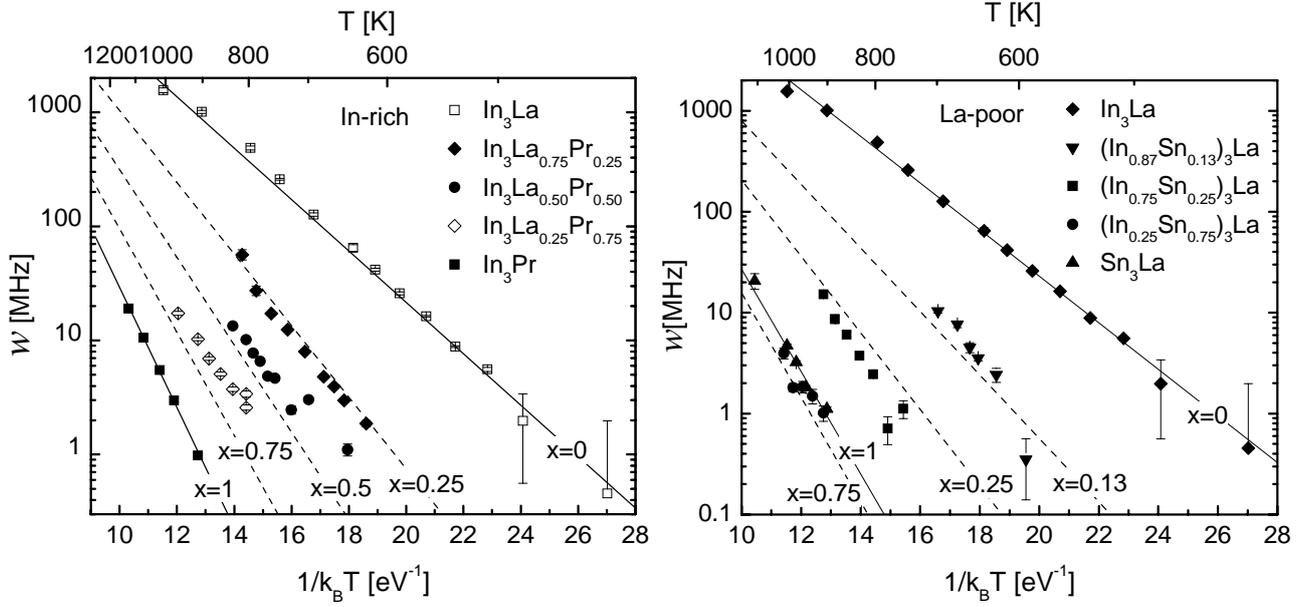

Fig. 2. Arrhenius plots of jump frequencies of $^{111}$Cd tracer probe atoms in pseudo-binary alloys In$_3$(La$_{1-x}$Pr$_x$), left, and (In$_{1-x}$Sn$_x$)$_3$La, right. Dashed lines are from models described in the text.

**Model to simulate composition dependences of jump frequencies**

A vacancy diffusion mechanism is assumed for both pseudo-binary systems, in which PAC probe atoms on the In or (In,Sn) sublattice exchange with a neighboring vacancy on the same sublattice. A model was developed to interpret the composition trends shown in Fig. 2 based on consideration of localized atomistic jumps. It is assumed that the mixed atoms are located at random on their respective sublattices, so that the probabilities of different local configurations are binomial.

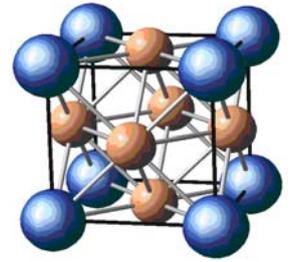

Fig. 3. L1$_2$ structure.

Fig. 3 shows the L1$_2$ crystal structure, with atoms on the (In,Sn) and (La,Pr) sublattices indicated by small, face-centered, and large corner spheres. For In$_3$(La$_{1-x}$Pr$_x$), diffusion takes place by jumps of probe atoms into vacant sites on the otherwise pure In-sublattice. Due to the differing size of Pr-solute atoms, changing lattice parameter, or other reasons, it is assumed that the activation enthalpy for a probe atom to jump into a vacancy is a function of the local number of Pr-atoms $n$ on the four R-atom sites that neighbor the vacancy (which can vary between 0 and 4). The activation enthalpy for jumping into a site with $n$ neighboring Pr-atoms is assumed to be some unknown function of $x$, $Q_n(x)$. An estimate of the mean activation enthalpy $\overline{Q}(x)$ can then be obtained as the average of $Q_n(x)$, weighted by the binomial probabilities for having $n$ solutes out of a total of $N$ local sites:

$$\overline{Q}(x) \equiv \sum_{n=0}^{N} Q_n(x) \binom{N}{n} x^n (1-x)^{N-n}, \qquad (2)$$

in which $N$ is the maximum number of relevant sites and $(N/n)$ are binomial coefficients. Measured end-member activation enthalpies obviously define the values $\overline{Q}(0) \equiv Q_0(0)$ and $\overline{Q}(1) \equiv Q_N(1)$.

For (In$_{1-x}$Sn$_x$)$_3$La, an analogous approach is taken. Here, the La-sublattice is full, and probe atoms jump from one site on the mixed (In,Sn) sublattice into a vacancy on the same sublattice. It is assumed that the activation enthalpy for a probe atom to jump into the vacancy is a function of the total number of Sn-atoms $n$ occupying the $N$=12 (In,Sn) sites that are near-neighbors to both the jumping atom and vacancy. Eq. 2 can again be used to calculate the mean activation enthalpy as a function of $x$, but with $N=12$ and a different set of functions $Q_n(x)$.

In order to allow for an explicit dependence of the activation enthalpies on composition, in addition to local configurations, the following linear dependence was assumed for $Q_n(x)$:

$$Q_n(x) = Q_0(0) + n\left(\frac{\overline{Q}(1) - \overline{Q}(0)}{N} - a(1-x)\right). \tag{3}$$

Here, $a$ defines the magnitude of the explicit composition dependences. Finally, in order to calculate the average activation enthalpy $\overline{Q}(x)$ for comparison with experiment, Eq. 3 is inserted in Eq. 2. Note that $a$ is the only adjustable model parameter for a given value of $N$. For each system, values of $a$ were adjusted to give the best overall simulation of data shown in Fig. 2. Best values found were $a= 0.0$ for $In_3(La_{1-x}Pr_x)$ and $a= +0.075$ eV for $(In_{1-x}Sn_x)_3La$. Dashed lines in Fig. 2 show results from the models for the nominal compositions $x$ indicated in the figure. As can be seen, agreement with experiment is reasonably good for both systems.

**Interpretation of experimental results using the model**. Fig. 4 shows composition dependences of the individual activation enthalpies $Q_n(x)$ for $In_3(La_{1-x}Pr_x)$, left, and $(In_{1-x}Sn_x)_3La$, right. For $In_3(La_{1-x}Pr_x)$, with $a=0$, the activation enthalpies $Q_n(x)$ are independent of $x$. For $(In_{1-x}Sn_x)_3La$, with $a=+0.075$ eV, $Q_n(x)$ are seen to decrease linearly as $x$ increases. Dark, red lines that cut across the lines for $Q_n(x)$ show mean activation enthalpies $\overline{Q}(x)$ calculated as a function of composition according to Eq. 2. As can be seen, the mean activation enthalpy $\overline{Q}(x)$ for $In_3(La_{1-x}Pr_x)$ increases linearly with $x$, with $d\overline{Q}(x)/dx \cong 0.63$ eV. The linear dependence implies that the jump frequency $w(x)$ scales with composition as $w(x)/w(0) \cong (w(1)/w(0))^x$. Thus, substitution of Pr for La has a benign effect on diffusion.

For $(In_{1-x}Sn_x)_3La$, Fig. 4 right, the mean activation enthalpy $\overline{Q}(x)$ increases rapidly for small $x$, with initial slope $d\overline{Q}(x)/dx \cong 1.6$ eV, but becomes essentially independent of $x$ and equal to the activation enthalpy of end-member phase $Sn_3La$ over the range $x\sim$ 0.6-1.0. Thus, there is a large reduction in jump frequency when Sn starts to substitute for In in $In_3La$, but essentially no change when In substitutes for Sn in $Sn_3La$. In effect, Sn-atoms in $In_3La$ act as "speed bumps" to reduce jump rates in their vicinity. To explain this behavior, we consider the atomic size and metal chemistry of impurities in the phases. Since $Sn_3La$ has only a slightly larger lattice parameter than $In_3La$, 0.477 vs. 0.473 nm, the strong reduction in jump rates does not appear to be caused by an oversized Sn impurity. Recalling that the diffusion tracer is the $^{111}$Cd daughter probe, one can ask how metal chemistry might affect diffusion in $In_3La$ containing small quantities of Sn, and diffusion in $Sn_3La$ containing small quantities of In. For this, we apply a naïve model based on nominal valences of sp-atoms. Relative to the perfect $In_3La$ structure, charges of an In-vacancy, Sn-impurity and Cd-impurity replacing an In-atom are -3, +1 and -1. Consequently, Cd-probes may become attracted to Sn-impurities, reducing their overall jump rate. Relative to the perfect $Sn_3La$ structure, charges of a Sn-vacancy, In-impurity and Cd-impurity replacing a Sn-atom are -4, -1 and -2. Thus, Cd should not be attracted to the indium impurities, leading to a small effect on jump rates for low In-concentrations. Such metal chemistry effects can be expected to be much smaller in $In_3(La_{1-x}Pr_x)$ because the valences of La and Pr are identical. Interestingly, the greater jump frequency of Cd in $In_3La$ than in $Sn_3La$ might be attributed in a general way to a more repulsive interaction between a Cd-probe and Sn-vacancy: (-2*-4= +8) than In-vacancy (-1*-3= +3).

Finally, recall that diffusion in end-members $In_3La$ and $Sn_3La$ of the mixed $(In_{1-x}Sn_x)_3La$ phases appear to be governed by different vacancies, respectively La-vacancies and Sn-vacancies [2,4]. It is perhaps not surprising that complex behavior would be observed as a function of composition.

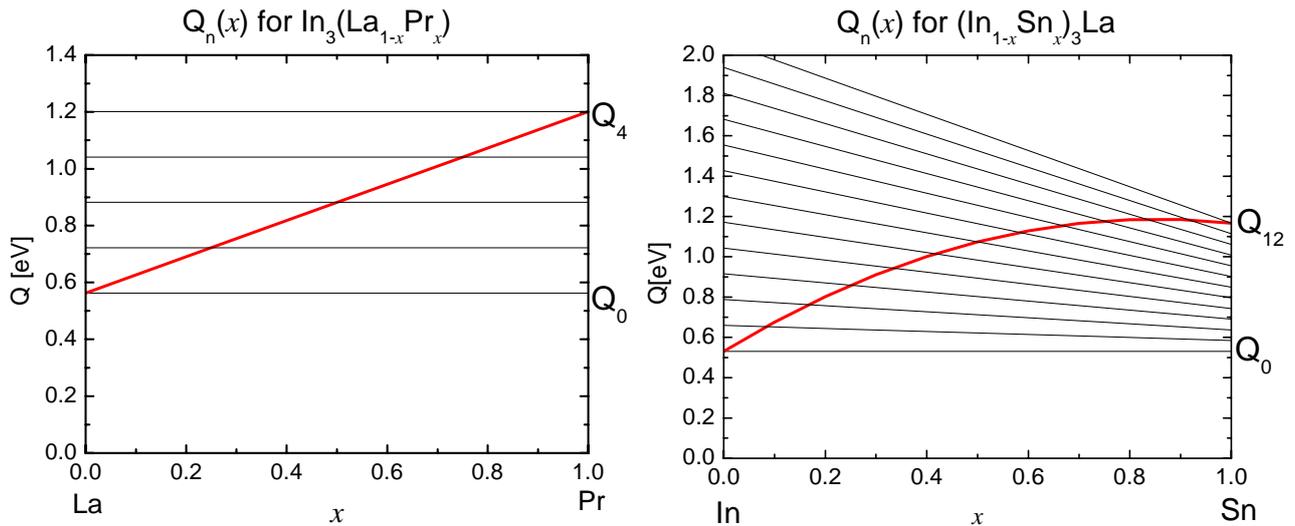

Fig. 4. Composition dependences of jump-frequency activation enthalpies modeled for probe atoms on In or (In,Sn) sublattices to jump into a vacancy on the same sublattice. Lines correspond to different local configurations of atoms.  <u>Left</u>:  $In_3(La_{1-x}Pr_x)$, with jumps into vacancies having 0-4 Pr-neighbors.  <u>Right</u>:  $(In_{1-x}Sn_x)_3La$, with jumps into vacancies having 0-12 Sn-neighbors.  Dark lines show the mean activation enthalpy as a function of composition.

**Summary and Acknowledgment**

Jump frequencies of $^{111}$Cd probe atoms were measured in two pseudo-binary alloy systems.  For $In_3(La_{1-x}Pr_x)$, activation enthalpies varied linearly with composition.  For $(In_{1-x}Sn_x)_3La$, activation enthalpies varied nonlinearly and nonmonotonically with composition.  The difference appears to be connected with whether atomic disorder was on the diffusion sublattice or an adjacent one, whether or not end-member phases have the same dominant vacancy diffusion mechanism, and/or whether there is a valence difference between the mixing atoms.

This work was supported in part by the National Science Foundation under grant DMR 09-04096 (Metals Program).